\documentclass[journal]{IEEEtran} 
\usepackage{graphicx}
\usepackage[cmex10]{amsmath}
\usepackage{algorithmicx}
\usepackage{algpseudocode}
\usepackage{algorithm}
\usepackage{multirow}
\usepackage{xcolor}
\usepackage{amsfonts}
\usepackage{bm}
\usepackage{verbatim}
\usepackage{cite}
\usepackage[normalem]{ulem}
\graphicspath{{fig/}} 

\ifodd 1
\newcommand{\congc}[1]{{\color{magenta}(Cong: #1)}} \else
\newcommand{\congc}[1]{} \fi

  

\newlength{\mywidth} \makeatletter \newif\ifhbonecolumn \@ifclasswith{IEEEtran}{onecolumn}{\hbonecolumntrue}{\hbonecolumnfalse} \makeatother
\ifhbonecolumn
\usepackage[top=1.07in, bottom=1.05in, left=1.05in, right=1.05in]{geometry} \fi

\begin{document}
\title{Hybrid Offline-Online Multi-Agent Decision Transformers for Wireless Resource Management}
\author{Yiming Zhang, Kun Yang, Cong Shen, Dongning Guo%
\thanks{This work was supported in part by the National Science Foundation under Grant Nos.~2132700, 2216970, and 2434044.}%
\thanks{Y. Zhang and D. Guo are with the Department of Electrical and Computer Engineering, Northwestern University, Evanston, IL 60208 USA (e-mail: yimingzhang2026@u.northwestern.edu; dguo@northwestern.edu).}%
\thanks{K. Yang and C. Shen are with the Department of Electrical and Computer Engineering, University of Virginia, Charlottesville, VA 22904 USA (e-mail: ky9tc@virginia.edu; cong@virginia.edu).}}


\maketitle
\thispagestyle{empty} 

\begin{abstract}
This paper develops a hybrid offline-online multi-agent reinforcement learning framework based on decision transformers. The policy is first pretrained offline via supervised sequence modeling of trajectories generated by existing policies, providing a safe and sample-efficient initialization. It is then fine-tuned online using a hybrid objective that incorporates critic-guided gradients, enabling performance improvements beyond the offline policy. To facilitate stable offline-to-online transfer and effective multi-agent coordination, the framework incorporates return-weighted sampling, a critic conditioned on neighbors' actions, and neighborhood-correlated exploration. The approach is fully distributed: both training and execution rely only on local observations and limited information exchange among neighboring agents. Evaluations with dynamic traffic arrivals in two settings: (i) joint scheduling and power allocation and (ii) coordinated beamforming, show that the proposed method achieves quality-of-service (QoS) performance comparable to centralized methods. Moreover, when pretrained on lower-quality datasets, online fine-tuning is also observed to surpass the initial offline policy. These results demonstrate a promising learning-based alternative for wireless resource management.
\end{abstract}


\section{Introduction}
\label{sec:Intro}

Interference management is a key determinant of quality of service (QoS) in dense wireless networks. Model-based optimization methods, such as weighted minimum mean-squared error (WMMSE)~\cite{shi2011iteratively} and fractional programming (FP)~\cite{shen2018fractional}, achieve near-optimal performance, but their dependence on global channel state information (CSI) and centralized computation becomes prohibitive in practical deployments. In contrast, widely deployed heuristics such as greedy max-SINR selection~\cite{3gpp2024ts38213} rely only on local measurements and simple decision rules, often at the cost of substantial performance degradation. This motivates the central question of this work: \emph{can a distributed solution operate efficiently using only local information while achieving QoS performance comparable to that of centralized optimization methods?}

To address this challenge, we introduce a hybrid offline-online multi-agent reinforcement learning (MARL) framework and evaluate it on wireless resource allocation problems. As a model-free framework, MARL naturally handles distributed systems where agents make sequential decisions under partial observability --- a setting common to wireless networks, traffic control, robotics, and smart grids. In the context of our wireless resource allocation application, MARL's architecture aligns perfectly with the problem structure: interference coupling creates distributed decision-making interactions among transmitters, network dynamics require continual adaptation, and devices typically have access only to local measurements. Consequently, online MARL has been studied for power control~\cite{nasir2019multi}, joint resource allocation~\cite{tan2020deep,nasir2020deep,khan2020centralized,zhang2025multi}, and user scheduling~\cite{yang2022asilomar,tefera2023deep,ge2023deep,zhang2024traffic}. However, online learning presents a critical deployment barrier: the random exploration required in the early learning stages can lead to poor QoS, while learning an effective policy from scratch can be highly sample-inefficient.

\emph{Offline reinforcement learning} provides a natural complement to online learning~\cite{yang2024offline, yang2024dyspan,zhang2025madt,eldeeb2026offline}. Instead of relying on potentially risky and inefficient online exploration, offline RL trains on pre-collected datasets. For example, the offline method in our prior work~\cite{zhang2025madt} succeeds in both sum-rate maximization and delay minimization in the joint scheduling and power allocation task and exhibits these advantages. In particular, high-performing but impractical centralized algorithms can generate synthetic trajectories from which their behavior is distilled into deployable distributed policies. The resulting agents can therefore benefit from mature optimization techniques without requiring centralized computation during deployment. Offline pretraining also offers improved safety, stability, and convergence over learning online from scratch.

A purely offline approach, however, has important limitations. A policy may degrade under distributional shift when deployment conditions differ from those represented in the training data. More fundamentally, its performance is constrained by dataset quality: policies trained on low-reward trajectories generally struggle to achieve expert-level performance~\cite{nakamoto2023cal}. We therefore propose a hybrid offline-online framework that combines safe and efficient offline pretraining with online fine-tuning. The pretrained policy provides a strong initialization, while subsequent interaction enables adaptation to deployment conditions and improvement beyond the data-generating policy, even when only low-quality offline data are available.

A key design choice in this hybrid framework is the policy architecture. Our prior work~\cite{zhang2024traffic} showed that sequential information and decision is crucial for dynamic wireless resource allocation, as CSI and queues evolve over time. Transformers~\cite{vaswani2017attention} have emerged as a leading architecture for sequence modeling, and the decision transformer~\cite{chen2021decision} brings this capability to RL by casting policy learning as return-conditioned sequence modeling. This formulation supports supervised pretraining on offline data and is therefore well suited to our hybrid framework. Our prior conference work~\cite{zhang2025madt} introduced the first multi-agent decision transformer (MADT) implementation for wireless resource management but considered only offline pretraining. Concurrent work~\cite{zhang2025decision} employs hybrid offline-online training of a single-agent decision transformer and targets sum-rate maximization in an unmanned aerial vehicle scenario, whereas we develop a fully distributed multi-agent framework for QoS optimization.

The main contributions of this paper are as follows:
\begin{itemize}
    \item We develop a hybrid offline-online MARL framework based on MADTs. Offline pretraining uses trajectories from existing policies to provide a safe and sample-efficient initialization. Online fine-tuning then combines supervised loss with critic-guided gradients, enabling improvement beyond the original offline policy.

    \item We address three challenges of fine-tuning under partial observability: nonstationarity from co-evolving neighbors, credit assignment with unobserved concurrent actions, and coordinated exploration. Our solutions consist of return-weighted anchored replay with critic warm-up, a critic conditioned on neighbors’ actions, and neighborhood-correlated exploration, all using only local observations and neighborhood information exchange.

    \item The framework is fully distributed and agnostic to the controlled resource: new resource-management problems require redefining only the per-agent observation, action, and reward. We demonstrate this generality on two instantiations with distinct actions and network deployment: joint scheduling and power allocation among interfering links and coordinated beamforming in a sectored multi-cell cluster. Both adopt packet delay as the QoS metric because it better reflects user experience than sum-rate in realistic traffic demand lighter than the network capacity. In both settings, simulation results show that the learned policies achieve QoS performance comparable to genie-aided centralized methods, and that online fine-tuning is able to improve upon the very policies that generated the offline data.
\end{itemize}


The remainder of this paper is organized as follows. We describe the general online-offline framework in Sec.~\ref{sec: MARL Framework}. Sec.~\ref{sec: System Model} introduces the wireless network scenarios, QoS, and MARL design. Sec.~\ref{sec: simulation results} presents the simulation setup and numerical results. Concluding remarks are given in Sec.~\ref{sec:Con}.

\section{MARL Framework}
\label{sec: MARL Framework}
\subsection{MARL Framework}

We consider a multi-agent system in which $K$ agents, indexed by $k \in \mathcal{K} = \{1, \ldots, K\}$, interacting with the environment over episodes of length-$\mathcal{T}$ time slots. The following is defined for each agent $k\in\mathcal{K}$ and time slot $t\in\{1,\dots,\mathcal{T}\}$:
\begin{itemize}
    \item The environment is described by a global state $\mathbf{s}^{(t)} \in \mathcal{S}$, where $\mathcal{S}$ denotes the state space.
    \item Agent $k$ observes only a partial view $O_k^{(t)} \in \Omega_k$ of the global state, where $\Omega_k$ is agent $k$'s observation space.
    \item Agent $k$ forms a belief, or history, $h_k^{(t)}$
    based on its local observation $O_k^{(t)}$, the observations shared by the agent's neighbors, and the history of these observations. The agent then takes an action $a_k^{(t)} \in \mathcal{A}_k$ based on the belief, where $\mathcal{A}_k$ denotes agent $k$'s action space.
    \item The agents' actions collectively form the joint action $\mathbf{a}^{(t)} = \Big(a_1^{(t)}, \ldots, a_K^{(t)}\Big) \in \mathcal{A}$, where $\mathcal{A} = \mathcal{A}_1 \times \dots \times \mathcal{A}_K$.
    \item The environment transitions to the next state according to the Markovian transition kernel
    $\mathcal{P}\big( \mathbf{s}^{(t+1)} \,\big|\, \mathbf{s}^{(t)},\mathbf{a}^{(t)} \big)$.
    \item Agent $k$ receives an individual reward denoted as $r_k^{(t)} = R_k\big(\mathbf{s}^{(t)}, \mathbf{a}^{(t)}, \mathbf{s}^{(t+1)}\big)$.
\end{itemize}

We model the system as a decentralized partially observable Markov decision process with individual rewards (Dec-POMDP-IR)~\cite{zhang2025multi}, which replaces the single team reward in a Dec-POMDP~\cite{oliehoek2016concise} by the $K$ individual reward functions $R_1,\dots,R_K$ introduced above. In addition, each agent's observation is a deterministic function of the global state rather than a stochastic one.

\begin{figure}
\centering
\includegraphics[width=.8\mywidth]{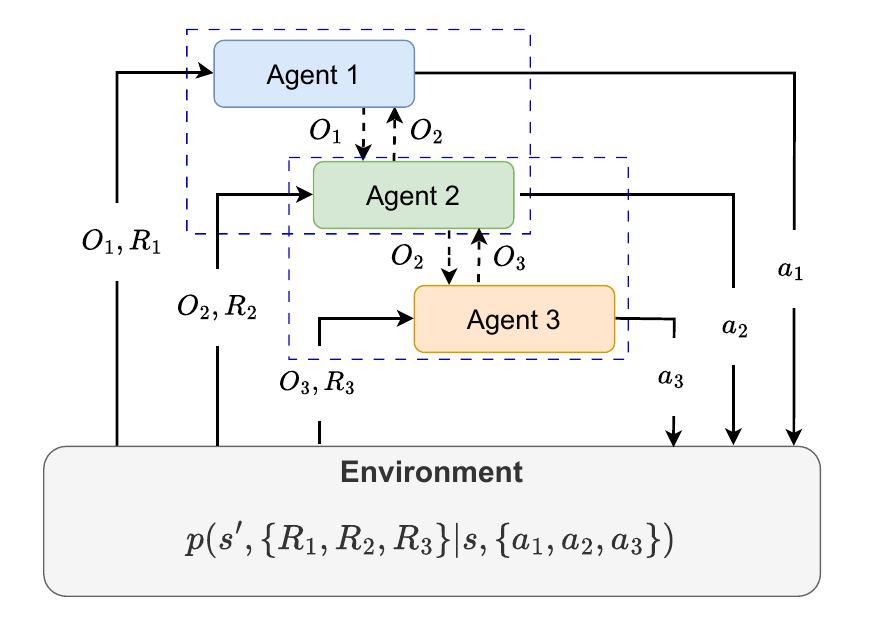}
\caption{An example of Dec-POMDP-IR.}
\label{fig: Dec-POMDP-IR}
\end{figure}

We define a neighbor relation between agents: two agents are said to be neighbors if 
they share their local observations with each other. Let $l_k$ denote the number of neighbors of agent $k$, and $\nu_{k,1},\dots,\nu_{k,l_k} \in \mathcal{K}$ denote their indices. Agent $k$'s neighborhood $\mathcal{D}(k)= (k,\nu_{k,1},\dots,\nu_{k,l_k})$ comprises the agent itself and its neighbors. Fig.~\ref{fig: Dec-POMDP-IR} illustrates this framework with three agents: agents~1 and~2 are neighbors, as are agents~2 and~3, while agents~1 and~3 are not neighbors and do not directly share information.

Agent $k$'s {\em policy}, $\pi_k$, represents a conditional probability distribution from which it samples its action $a_k$. Agent $k$'s learning goal is to compute a good policy to maximize its own expected return accumulated over an episode of $\mathcal{T}$ slots:
\begin{align} \label{eq: agent objective}
    \mathbb{E}_{\pi} \left[\sum_{t=1}^{\mathcal{T}} \gamma^{t-1}
    R_k\left(\mathbf{s}^{(t)}, \mathbf{a}^{(t)}, \mathbf{s}^{(t+1)}\right)\right],
\end{align}
where $\pi$ denotes the joint policy and $\gamma\in[0,1]$ is the discount factor.

\subsection{Offline Pretraining with Decision Transformers}
\label{subsec: offline pretraining} 

We adopt a decision transformer as the policy network in the multi-agent setting. Let the superscript $(s:t)$ denote a sequence spanning slots from $s$ to $t$. For agent $k$, the local aggregate information $X_k^{(t)} = \big( O_k^{(t)}, O_{\nu_{k,1}}^{(t)}, \dots, O_{\nu_{k,l_k}}^{(t)} \big)$ collects its own and its neighbors' observations. The return-to-go $\hat{R}_k^{(t)} = \sum_{l = t}^{\mathcal{T}} r_k^{(l)}$ is the tail of the finite-horizon objective~\eqref{eq: agent objective} with $\gamma = 1$. The belief $h_k^{(t)} = \big(X_k^{(t-L+1:t)}, a_k^{(t-L+1:t-1)}, \hat{R}_k^{(t-L+1:t)}\big)$ collects the length-$L$ history of aggregate information, actions, and returns-to-go. The transformer maps the belief to an action distribution, $a_k^{(t)} \sim \pi_k\big(\cdot \,\big|\, h_k^{(t)}\big)$, forming an auto-regressive model of order $L$. In addition to embeddings of the return-to-go, aggregate information, and action, the decision transformer uses a learned embedding of the within-episode slot index $t$ as positional encoding.

Offline pretraining offers two key advantages: i) safe initial deployment by imitating proven policies, and ii) improved training efficiency through curated data rather than online exploration. In~\cite{hu2024qvalue, zhang2025madt}, decision transformers trained offline converge to the performance level of the behavior policies that generated the data.

Offline pretraining consists of two phases. First, during data collection, a baseline method is executed over multiple simulated episodes. At each slot, we record the local observations, actions, and rewards available to each agent and compute the corresponding returns-to-go to construct the offline dataset. Second, all agents' trajectories {are used to train a shared} policy network $\pi_{\text{DT}}$ via supervised learning.
Let a length-$L$ trajectory window beginning at slot $t$ be $\tau^{(t)} = \big( \hat{R}_k^{(t:t+L-1)}, X_k^{(t:t+L-1)}, a_k^{(t:t+L-1)} \big)_{k \in \mathcal{K}}$. For each training step, a mini-batch $(\tau^{(l)})_{l\in \mathcal{B}}$ of such windows is sampled, and the policy is updated by minimizing the cross-entropy loss over all agents and slots:
\begin{align} \label{eq:offline loss function}
\mathcal{L}_{\text{CE}} ={}& -\frac{1}{|\mathcal{B}| K L} \sum_{k=1}^K \sum_{l\in \mathcal{B}} \sum_{t=l}^{l+L-1} \nonumber \\
& \log\pi_{\text{DT}}\left( a_k^{(t)} \,\middle|\, X_k^{(l:t)},\hat{R}_k^{(l:t)}, a_k^{(l:t-1)}\right).
\end{align}
A causal attention mask ensures that the prediction at slot $t$ conditions only on tokens up to $t$. At execution, the target return is initialized to the mean of the top-quartile episode returns of the offline dataset and decremented by the realized reward at each slot, prompting the transformer to reproduce the dataset's best behavior. The trained policy is executed independently by each agent.

\subsection{Online Fine-Tuning}
\label{subsec: online finetuning} 

Purely offline learning has two key limitations. First, high-quality policies may be unavailable or impractical for real-world data collection. For instance, FP and WMMSE require centralized coordination and real-time global CSI, so they can be emulated to synthesize data but cannot be deployed to collect real-world data. Second, deployment conditions may differ substantially from the training distribution, leading to out-of-distribution actions and compounding errors that hinder adaptation.

To mitigate distribution shift, the online decision transformer~\cite{zheng2022online} fine-tunes an offline-pretrained model through online interaction while retaining the auto-regressive supervised-learning paradigm. However, supervised learning alone cannot overcome suboptimal low-reward data because it provides no mechanism for improving beyond the collected trajectories.

We therefore augment the supervised objective with a policy-improvement term, allowing the policy to surpass rather than merely imitate the behavior in its training data. During online fine-tuning, the actor minimizes
\begin{align} \label{eq:online loss function}
\mathcal{L}_{\text{online}} = \mathcal{L}_{\text{CE}} + \lambda_{\text{RL}} \, \mathcal{L}_{\text{RL}} - \lambda_{\text{ent}} \, \bar{\mathcal{H}},
\end{align}
where $\mathcal{L}_{\text{CE}}$ is the cross-entropy loss in~\eqref{eq:offline loss function}, evaluated on the fine-tuning batch, which regularizes the policy toward proven behavior, $\mathcal{H}(\cdot)$ denotes the policy entropy, whose mini-batch average $\bar{\mathcal{H}} = \frac{1}{|\mathcal{B}|K}\sum_{k=1}^{K}\sum_{l\in\mathcal{B}} \mathcal{H}\big(\pi_{\text{DT}}(\cdot\,|\,h_k^{(l)})\big)$ encourages continued exploration, while the reinforcement-learning term $\mathcal{L}_{\text{RL}}$, defined shortly, drives improvement beyond the dataset. The hyperparameters $\lambda_{\text{RL}}$ and $\lambda_{\text{ent}}$ balance the three objectives.

The RL term requires estimating how much each action improves upon the current policy. A Monte-Carlo policy gradient could use the empirical return-to-go $\hat{R}_k^{(t)}$ for this purpose, but such estimates have high variance and do not bootstrap. Following the RL-as-fine-tuning approach of~\cite{yan2024reinforcement}, we instead learn twin critics, $Q_{\phi_1}$ and $Q_{\phi_2}$, shared by all agents, which maps an agent's aggregate information $X_k^{(t)}$ to one value per discrete action. This constitutes a discrete-action counterpart of TD3~\cite{fujimoto2018addressing}. Unlike the sequence-model actor, each critic is a lightweight, memoryless network that consumes only the current aggregate information, a choice observed in~\cite{yan2024reinforcement} to be substantially more stable than a transformer critic. As in TD3, the second critic is only used to reduce overestimation in the bootstrap target introduced below, whereas the actor is updated using the first critic. The resulting advantage of action $a$ is
\begin{align} \label{eq:advantage}
\begin{split}
A_{\phi_1}\!\left(X_k^{(t)}, a\right) &={} Q_{\phi_1}\!\left(X_k^{(t)}, a\right) \\
& \quad- \sum_{a' \in \mathcal{A}_k} \pi_{\text{DT}}\!\left(a' \,\middle|\, h_k^{(t)}\right) Q_{\phi_1}\!\left(X_k^{(t)}, a'\right).
\end{split}
\end{align}
The RL loss increases the likelihood of above-average actions:
\begin{align*}
\mathcal{L}_{\text{RL}} = -\frac{1}{|\mathcal{B}| K} \sum_{k=1}^{K} \sum_{l \in \mathcal{B}} \sum_{a \in \mathcal{A}_k} \pi_{\text{DT}}\!\left(a \,\middle|\, h_k^{(l)}\right) \frac{A_{\phi_1}\!\left(X_k^{(l)}, a\right)}{\sigma_A}
\end{align*}
where $\sigma_A$ is the standard deviation of the advantage, making the gradient scale insensitive to reward magnitude. Unlike hard filtering of positively-advantaged samples, this expectation-based objective yields a smoother update that reweights the entire action distribution.

The critics are trained by temporal-difference learning, i.e., by regressing $Q_{\phi_i}(X_k^{(t)}, a_k^{(t)})$ onto a bootstrap target formed from the reward and the estimated value of the next slot. Because a target computed directly from the networks being trained would change after every gradient step, regression onto it can be unstable~\cite{fujimoto2018addressing}. We therefore compute the target from quasi-static copies of the learned networks: target critics $\{Q_{\bar{\phi}_1}, Q_{\bar{\phi}_2}\}$ and a target actor $\bar{\pi}_{\text{DT}}$, which are held fixed within each update and slowly track their online counterparts via Polyak averaging, $\bar{\phi}_i \leftarrow \kappa\,\phi_i + (1-\kappa)\,\bar{\phi}_i$ with $\kappa \ll 1$ (likewise for $\bar{\pi}_{\text{DT}}$). The target evaluates the next slot by averaging, over the target actor's next-step action distribution, the element-wise minimum of the two target critics' predictions, which counteracts the overestimation bias of learned value functions~\cite{fujimoto2018addressing}:
\begin{align*} 
    y_k^{(t)} = r_k^{(t)} + \gamma_{\text{c}} \!\!\sum_{a' \in \mathcal{A}_k} \!\! \bar{\pi}_{\text{DT}}
    \!\big(a' \,|\, h_k^{(t+1)} \big)
    \min_{i \in \{1,2\}} Q_{\bar{\phi}_i}\!\big( X_k^{(t+1)}, a' \big)
\end{align*}
where $\gamma_{\text{c}} \in (0,1)$ is the bootstrapping discount of the critic. Both critics regress onto this target:
\begin{align} \label{eq:critic loss}
\mathcal{L}_{\text{critic}} = \frac{1}{|\mathcal{B}| K} \sum_{k=1}^{K} \sum_{l \in \mathcal{B}} \sum_{i \in \{1,2\}} \Big( Q_{\phi_i}\!\big(X_k^{(l)}, a_k^{(l)}\big) - y_k^{(l)} \Big)^2 .
\end{align}

Online fine-tuning proceeds iteratively. In each iteration, agents collect trajectories and append them to a replay buffer that retains the best offline trajectories. Mini-batches drawn from this buffer are used for several critic gradient steps minimizing~\eqref{eq:critic loss}, followed by actor updates minimizing~\eqref{eq:online loss function}, with target networks updated through Polyak averaging after each step. More frequent critic updates help maintain accurate advantage estimates~\cite{fujimoto2018addressing}.

\subsection{Stability and Coordination in Online Fine-Tuning}
\label{subsec: stability coordination} 

Online fine-tuning in a partially observable multi-agent environment introduces challenges absent from the single-agent setting. Each agent's reward depends on its neighbors' concurrent actions. Meanwhile, co-evolving neighbor policies create non-stationary learning targets, and improvements often require \emph{coordinated} behavior across a neighborhood, which independent exploration is unlikely to discover. The offline-to-online transition is also delicate because an untrained critic can destabilize a capable pretrained policy. We employ several mechanisms to address these challenges.

\textit{Stabilizing the offline-to-online transition.} We first warm up the critics with the actor frozen, and ramp $\lambda_{\text{RL}}$ from zero over the first few iterations, so the RL gradient is trusted only once the critic is informative. The replay buffer keeps a fixed \emph{anchor} set of the highest-return offline trajectories; each mini-batch mixes anchor and online data in fixed proportion, sampling trajectories with probability increasing in their standardized returns, so the supervised term imitates the best behavior seen so far rather than the latest exploratory episodes. Finally, the return target $\hat{R}_{\text{tar}}$ used for data collection is the running maximum of observed episode returns, preventing poor exploratory episodes from lowering the target.

\textit{Conditioning the critic on neighbors' actions.} Agent $k$'s reward depends on its neighborhood's joint action, yet $X_k^{(t)}$ carries no information about the neighbors' \emph{concurrent} actions. We condition the critic on the neighbors' intended actions at training time, $Q_{\phi}\big( X_k^{(t)}, a_k^{(t)}; \{a_j^{(t)}\}_{j \in \mathcal{D}(k) \setminus \{k\}} \big)$. This allows the critic to evaluate an action in the context of the neighbors’ behavior.

\textit{Temporally-extended, neighborhood-correlated exploration.} Coordinated actions may require neighboring agents select compatible actions over several consecutive slots. Independent $\epsilon$-greedy exploration lacks coordination while isolated deviation yields noisy learning signals for the critic because rewards depend on neighborhood joint actions. We therefore correlate exploration across both agents and time. In each slot, an agent $k$ is selected as the exploration center with probability $\epsilon$ and sends a one-bit trigger to its neighbors. Each agent in $\mathcal{D}(k)$ then independently samples an exploratory action and holds it for a fixed number of slots, while agents outside the neighborhood continue to follow the policy. An agent still committed to an earlier exploratory action retains that action. Each trigger thus perturbs the joint action over one complete neighborhood --- precisely the configuration on which the center agent's reward and critic depend --- without requiring common randomness or centralized coordination. This mechanism extends $\epsilon_z$-greedy exploration~\cite{dabney2020temporally} to the multi-agent setting.

The latter two mechanisms trade implementation complexity for coordination capability. They are particularly useful in large or tightly coupled systems, where rewards strongly depend on neighbors’ concurrent actions. In smaller or weakly coupled systems, a per-agent critic and standard action sampling may suffice.

\section{Wireless Model and Problem Formulation}
\label{sec: System Model}

\begin{figure*}
    \centering
    \includegraphics[width=\textwidth]{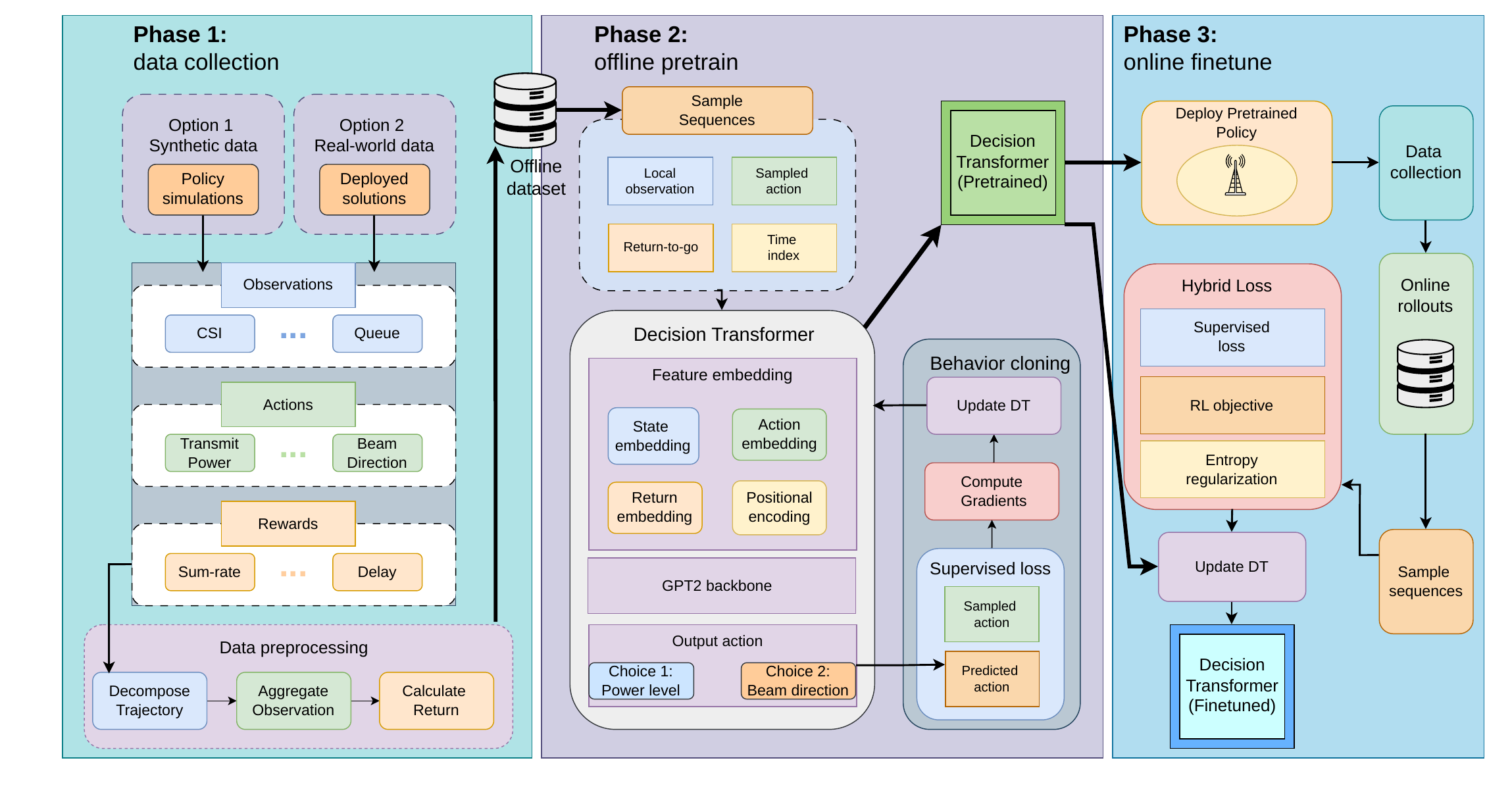}
    \caption{Three-phase hybrid offline-online MADT framework for wireless resource allocation.} 
    \label{fig:training workflow}
\end{figure*}

We instantiate the MARL framework in two wireless resource-allocation scenarios. Fig.~\ref{fig:training workflow} summarizes the three-phase training pipeline of Sec.~\ref{sec: MARL Framework}: (i) 
an offline dataset is either synthesized by simulating some existing policies or collected in a deployed system; (ii) offline pretraining employs supervised learning on this dataset to establish a safe initial decision-transformer policy; and (iii) online fine-tuning refines the policy through environmental interaction using the hybrid loss, enabling improvement beyond the quality of the original dataset while preserving distributed execution.

\subsection{Joint Scheduling and Power Allocation}
\label{subsec: Power Allocation}

\subsubsection{System Model}

\begin{figure} 
    \centering
    \includegraphics[width=.9\mywidth]{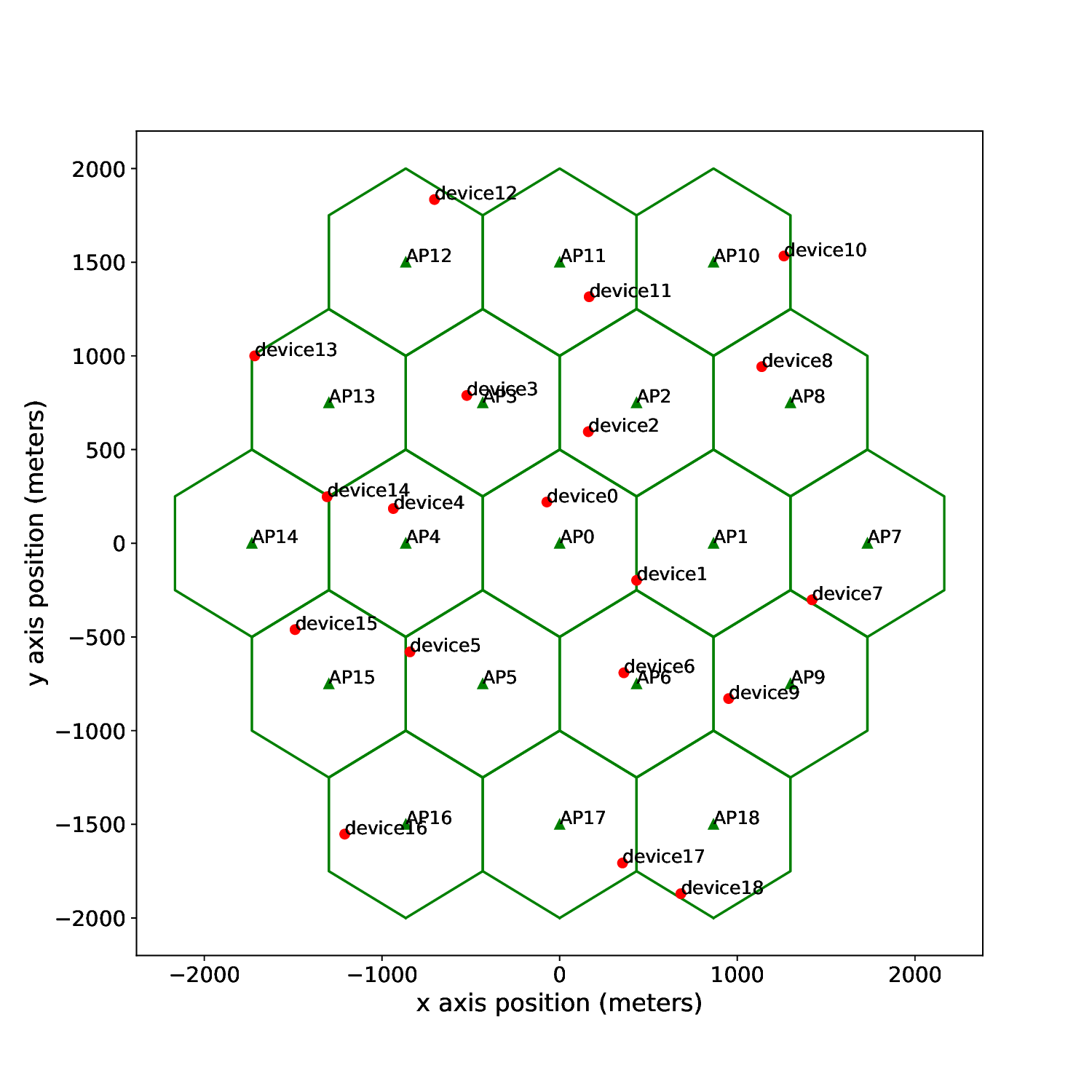}
    \caption{Nineteen hexagonal cells, each serving one active device.}
    \label{fig:deployment}
\end{figure}

We consider downlink in a network of $K$ transmitter-receiver pairs, representing a single channel or subchannel in a mobile ad hoc network or a cellular network in which each access point (AP) serves one active device. An example is shown in Fig.~\ref{fig:deployment}. Each link is treated as an agent. Neighboring agents communicate through low-latency interfaces, such as the Xn interface between gNBs~\cite{3gpp38420}. The associated fiber backhaul latency is typically~$1$~ms or less~\cite{jungnickel2013backhaul}, well below the slot duration $T = 20$~ms considered here.

Each transmitter and receiver has a single antenna. Under flat fading, the channel power gain from transmitter $i$ to receiver $j$ in slot $t$ is
\begin{align} \label{eq: downlink channel gain}
    g_{i\rightarrow j}^{(t)}=\alpha_{i\rightarrow j} \big|\beta_{i\rightarrow j}^{(t)}\big|^{2}, \quad t=1,2, \ldots
\end{align}
where $\alpha_{i\rightarrow j} \geq 0$ is the large-scale path loss component, which remains constant over many slots, and $\beta_{i\rightarrow j}^{(t)}$ is the small-scale Rayleigh fading coefficient. To capture temporal channel correlation, we model the latter as a first-order complex Gauss-Markov process:
\begin{align} \label{eq: small-scale fading}
    \beta_{i\rightarrow j}^{(t)}=\rho \beta_{i\rightarrow j}^{(t-1)}+\sqrt{1-\rho^{2}} e_{i\rightarrow j}^{(t)}
\end{align}
where $\rho \in [0,1]$ is the temporal correlation coefficient, and $\big( \beta_{i\rightarrow j}^{(0)}, e_{i\rightarrow j}^{(1)}, e_{i\rightarrow j}^{(2)}, \ldots \big)$ are independent and identically distributed circularly symmetric complex Gaussian random variables with unit variance.

Agent neighborhoods are defined by interference coupling. Specifically, agent $j$ belongs to agent $i$'s neighborhood if the relative path loss $\alpha_{j \rightarrow i} / \alpha_{i \rightarrow i}$ exceeds a predefined threshold, indicating nonnegligible interference from transmitter $j$ to receiver $i$. In a hexagonal deployment, we assume each agent typically has at most six neighbors.

Let $p_{k}^{(t)}$ denote the transmit power of transmitter $k$ in slot $t$, and let $\bm{p}^{(t)} = \big( p_{1}^{(t)}, \ldots, p_{K}^{(t)}\big)$ denote the global power-allocation vector. Assuming additive white Gaussian noise with variance $\sigma^2$ at each receiver, the spectral efficiency of link $k$ is
\begin{align} \label{eq: spectral efficiency}
\mathcal{C}_{k}^{(t)} \left(\bm{p}^{(t)}\right)= \log_2\left(1 + \frac{g_{k\rightarrow k}^{(t)} p_{k}^{(t)}}{\sum_{j \in \mathcal{K}, j \neq k} g_{j\rightarrow k}^{(t)} p_{j}^{(t)}+\sigma^{2}}\right)
\end{align}

We consider a traffic-driven system in which each link maintains a first-in, first-out (FIFO) queue. Let $\zeta_k^{(t)}$ denote the number of packets arriving at link $k$ at the beginning of slot $t$, and let $M$ denote the packet size in bits. Given bandwidth $W$, the queue length $q_k^{(t)}$, measured in bits at the end of the slot, evolves as
\begin{align} \label{eq: traffic dynamic}
    q_{k}^{(t)}
    =\max \left(0, \, q_{k}^{(t-1)} + \zeta_{k}^{(t)} M - \mathcal{C}_{k}^{(t)} W T \right) .
\end{align}

\subsubsection{Agent Design}
At the beginning of slot $t$, agent $k$'s local observation $O_k^{(t)}$ comprises:
\begin{itemize}
    \item $g_{k\rightarrow k}^{(t)}$: the current direct channel gain,
    \item $p^{(t-1)}_{k}$: agent $k$'s previous action,
    \item $q_{k}^{(t-1)}+\zeta_{k}^{(t)} M$: the queue backlog after new arrivals,
    \item $\sum_{j \in \mathcal{K}, j \neq k}g_{j\rightarrow k}^{(t-1)}~p^{(t-1)}_{j}+\sigma^{2}$: the total interference-plus-noise power,
    \item $\mathcal{C}_{k}^{(t-1)}$: spectral efficiency of link $k$ computed from~\eqref{eq: spectral efficiency}.
\end{itemize}
The last two quantities result from the previous slot's transmissions and are therefore available with a one-slot delay. Through neighborhood information exchange, agent $k$ obtains the aggregate information $X_k^{(t)} = \left\{ O_k^{(t)}, O_{\nu_{k,1}}^{(t)}, \dots, O_{\nu_{k,l_k}}^{(t)} \right\}$ before decision making in slot $t$.

Based on $X_k^{(t)}$, agent $k$ decides whether to transmit and, if so, at what power. It selects $p_{k}^{(t)}$ from a discrete action set comprising zero, representing no transmission, and power levels equally spaced in dB between the minimum nonzero power $\underline{P}$ and the maximum power $\overline{P}$:
\begin{align} \label{eq: action space}
    p_{k}^{(t)} &\in \mathcal{A}_k = \Big\{0, \, \underline{P}, \, \underline{P}
(\overline{P}/\underline{P})
^{\frac{1}{\left|\mathcal{A}_k\right|-2}}, \ldots, \, \overline{P}\Big\} .
\end{align}

We use queue length as a surrogate QoS objective because, by Little’s law, shorter time-averaged queues corresponds to shorter delays. The utility of agent $k$ is therefore $u_{k}^{(t)} = -q_{k}^{(t)}$. To promote coordination, agent $k$'s reward also includes the utilities of its neighbors. This discourages aggressive scheduling and transmission that reduces one's own backlog at the cost of greater interference and larger neighboring queues. Specifically,
\begin{align} \label{eq:reward function}
    { r_k^{(t)}
    =\sum_{i \in \mathcal{D}(k)} u_i^{(t)}
    = -\sum_{i \in \mathcal{D}(k)} q_i^{(t)}.}
\end{align}

\subsection{Beamforming}
\label{subsec: Beamforming}

\subsubsection{Network Model}
\begin{figure} 
    \centering
    \includegraphics[width=\mywidth]{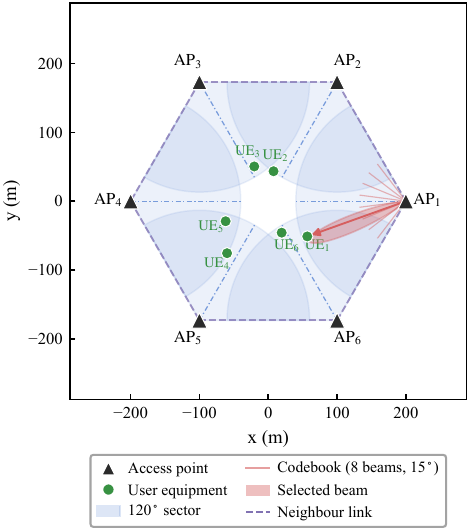}
    \caption{Beamforming deployment: six APs with inward-facing $120^\circ$ sectors form one cluster of mutually-interfering agents.}
    \label{fig:beamforming_deployment}
\end{figure}

Our second instantiation considers downlink beamforming in a sectored multi-cell network. Each site is partitioned into three $120^\circ$ sectors, with inter-cell interference dominated by sectors directly facing one another. We therefore consider a cluster of $K$ APs arranged on a ring with inward-facing sectors. Each AP serves one active device over a shared frequency band and is regarded as one agent. Its two adjacent APs are regarded as its neighbors, with which it exchanges observations as described in Sec.~\ref{sec: MARL Framework}. In each slot, an AP either makes no transmission or selects one beam from a codebook and transmits at a fixed power. Each device moves randomly across slots, producing temporally correlated channels. Fig.~\ref{fig:beamforming_deployment} depicts an example.

Because the sectors face one another, their beam selections are coupled: a beam aimed at one's own device also leaks power into other cells. Each AP must therefore balance its own rate against the interference it inflicts on others. As in the joint scheduling and power allocation setting, we consider traffic-driven delay-minimization. Poisson packet arrivals accumulate in a FIFO queue at each AP, which is served at the realized rate according to the queueing dynamics in~\eqref{eq: traffic dynamic}.

\subsubsection{Agent Design}
Agent $k$ stays silent or selects one of $B$ beams spanning its sector, i.e.,
$ a_k^{(t)} \in \mathcal{A}_k = \{0, 1, \ldots, B\}$
where $0$ corresponds to silence. Let $G(a,\theta)$ denote the gain of beam $a$ toward angle $\theta$, with $G(0,\theta)=0$. Let $\theta_{j\to k}^{(t)}$ denote the angle from AP $j$ to user $k$, $d_{j\to k}^{(t)}$ their distance, $\ell(\cdot)$ the path loss as a function of distance, and $P$ the transmit power. The spectral efficiency of user $k$ is
\begin{align} \label{eq: beam spectral efficiency}
\mathcal{C}_k^{(t)} = \log_2\!\left(1 + \frac{P\, G\!\left(a_k^{(t)}, \theta_{k\to k}^{(t)}\right) / \ell\!\left(d_{k\to k}^{(t)}\right)}{I_k^{(t)}
+ \sigma^2}\right)
\end{align}
where
$I_{j\to k}^{(t)} = P\, G\!\left(a_j^{(t)}, \theta_{j\to k}^{(t)}\right) / \ell\!\left(d_{j\to k}^{(t)}\right) $
denotes the interference that AP $j$'s transmission places on user $k$,
and $ I_k^{(t)} = \sum_{j \neq k} I_{j\to k}^{(t)} $
denotes the aggregate interference.

The beam gains toward the served device are measured at the beginning of each slot. Quantities resulting from transmission, including the realized spectral efficiency and received interference, become available after the slot and therefore carry a one-slot delay. Agent $k$'s local observation $O_k^{(t)}$ comprises the direct beam gains $G\big(b, \theta_{k\to k}^{(t)}\big)$, $b=1,\dots,B$, distance $d_{k\to k}^{(t)}$, queue backlog $q_{k}^{(t-1)}+\zeta_{k}^{(t)} M$, the previous action $a_k^{(t-1)}$, the previous realized spectral efficiency $\mathcal{C}_{k}^{(t-1)}$, the previous aggregate interference $I_k^{(t-1)}$, and the previous per-source interference $I_{\nu\to k}^{(t-1)}$ of each neighbor, estimated at user $k$'s receiver and fed back to AP $k$.

In this beamforming setting, 
agent $\nu$ shares only
\begin{align} \label{eq: neighbor message}
m_{\nu}^{(t)} = \big( \mathcal{C}_{\nu}^{(t-1)},\; a_{\nu}^{(t-1)},\; q_{\nu}^{(t-1)}+\zeta_{\nu}^{(t)} M \big) \subset O_{\nu}^{(t)}
\end{align}
with its neighbors. Assuming agent $k$ has two neighbors, i.e., $\mathcal{D}(k)=\{k, \nu_{k,1},\nu_{k,2}\}$, its aggregate information is 
    $X_k^{(t)} = \big( O_k^{(t)},\; m_{\nu_{k,1}}^{(t)},\; m_{\nu_{k,2}}^{(t)} \big)$.
The utility and the reward follow Sec.~\ref{subsec: Power Allocation}, i.e., 
$u_k^{(t)} = -q_k^{(t)}$, and $r_k^{(t)} 
= -\sum_{i \in \mathcal{D}(k)} q_i^{(t)}$.

\section{Performance Evaluation}
\label{sec: simulation results}

We evaluate the hybrid-RL framework on both settings introduced in Sec.~\ref{sec: System Model}. Throughout this section, the joint scheduling and power allocation task of Sec.~\ref{subsec: Power Allocation} is referred to simply as power allocation for brevity.

\subsection{Power Allocation}

\subsubsection{Simulation setup}

We consider a network of $19$ devices in $19$ hexagonal cells, as illustrated in Fig.~\ref{fig:deployment}. The channel, traffic, and action-space models follow Sec.~\ref{sec: System Model}, with the parameters listed in Table~\ref{tab:sim_params}; an agent's neighborhood is determined by the interference-coupling rule defined therein. Also, each agent digests its neighbors' shared observations by local index (its first neighbor, second neighbor, and so on), so the shared policy is agnostic to any global indexing of agents or links.

\begin{table}[t]
    \centering
    \caption{Simulation Parameters of the Two Instantiations}
    \label{tab:sim_params}
    \footnotesize
    \setlength{\tabcolsep}{3pt}
    \begin{tabular}{l|c|c}
        \hline
        \textbf{Parameter} & \textbf{Power allocation} & \textbf{Beamforming} \\
        \hline
        Cell radius & 500 m & 160 m \\
        Path loss ($d$ in km) & $128.1 + 37.6\log_{10} d$ dB & exponent $=2$ \\
        Transmit power & $(P_{\min}, \overline{P})=(0,23)$ dBm & fixed $P$ \\
        Noise power & $-114$ dBm & 10 dB cell-edge SNR \\
        Action space $\mathcal{A}_k$ & 7 power levels & 8 beams $+$ silent \\
        Channel variation & $f_d = 10$ Hz, $\rho = 0.64$ & user random walk \\
        Arrival rate $\lambda$ & 25 packets/s & 10 packets/s \\
        \hline
        \multicolumn{3}{l}{Common: $W = 10$ MHz, $T = 20$ ms, $M = 500$ kbits, 
        $\mathcal{T} = 1000$ slots.} \\
        \hline
    \end{tabular}
\end{table}

We compare the learned policies against the following baselines:
\begin{itemize}
    \item \textbf{Full Power}: all APs transmit at maximum power;
    \item \textbf{Random Power}: Each AP selects its power uniformly;
    \item \textbf{ITLinQ}~\cite{naderializadeh2014itlinq}: a \emph{distributed} scheduling method inspired by information theory, realizable with minor adaptation;
    \item \textbf{ideal WMMSE}~\cite{shi2011iteratively}: centralized WMMSE with \emph{instantaneous} global CSI, included as a genie-aided benchmark;
    \item \textbf{delayed WMMSE}: centralized WMMSE with one-slot-\emph{delayed} global CSI. This is realizable but unscalable.
\end{itemize}

Offline datasets are generated by two behavior policies of different quality: an \emph{expert} dataset produced by ideal WMMSE, and a \emph{medium} dataset produced by delayed WMMSE. Each dataset contains $40$ episodes ($40{,}000$ transitions) collected under independent channel and traffic realizations. 

Online fine-tuning then performs $200$ iterations, each of which collects one exploratory episode under a fresh channel/traffic realization and applies the critic and actor updates of Sec.~\ref{subsec: online finetuning}, i.e., an online interaction budget of only $200$ episodes. 
All reported evaluations are rollouts on a common held-out channel/traffic realization, whose seed is disjoint from those of the offline dataset and the online exploration episodes, so that no evaluation data is seen during training,
and every stage and baseline is scored on the same episode for a controlled comparison.


\subsubsection{Training dynamics and convergence}

\begin{figure*} 
    \centering
    \includegraphics[width=.92\textwidth]{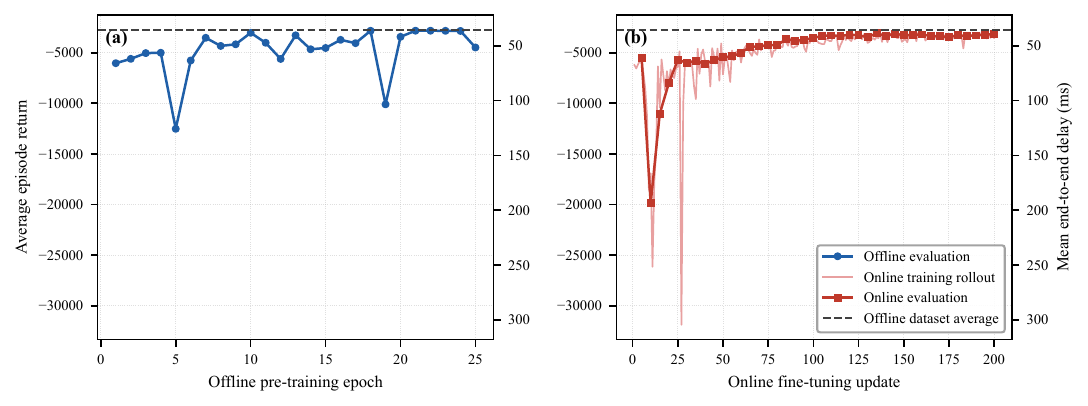}
    \caption{Learning curves when pretraining on the \emph{expert} (ideal WMMSE) dataset: (a) offline pretraining and (b) online fine-tuning. The dashed line marks the average episode return of the offline dataset. Each curve is read against both vertical axes: average episode return (left) and the corresponding mean end-to-end delay (right, inverted so that smaller delay is at the top).}
    \label{fig:lc_expert}
\end{figure*}

\begin{figure*} 
    \centering
    \includegraphics[width=.92\textwidth]{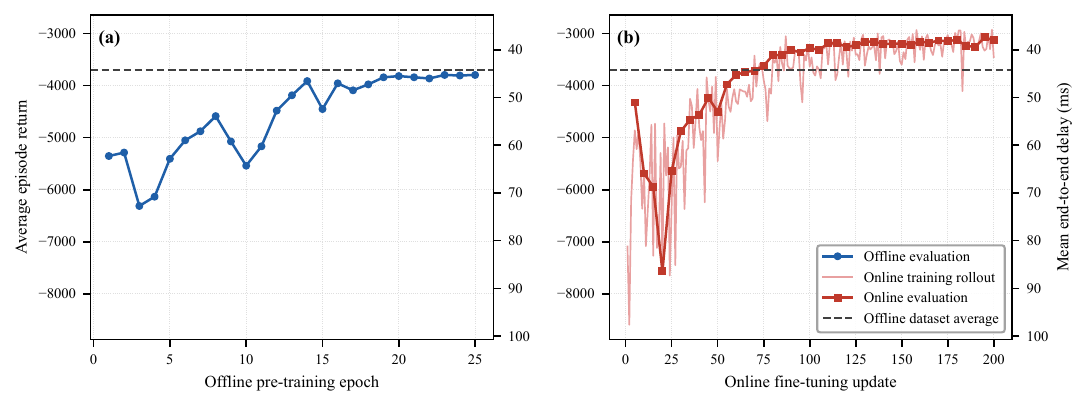}
    \caption[Learning curves when pretraining on the \emph{medium} (delayed WMMSE) dataset: (a) offline pretraining and (b) online fine-tuning. Axes as in Fig.~\ref{fig:lc_expert}. After the initial exploration phase, the fine-tuned policy surpasses the average return of the offline dataset.]{Learning curves when pretraining on the \emph{medium} (delayed WMMSE) dataset: (a) offline pretraining and (b) online fine-tuning. Axes as in Fig.~\ref{fig:lc_expert}. After the initial exploration phase, the fine-tuned policy surpasses the average return of the offline dataset} 
    \label{fig:lc_medium}
\end{figure*}

Figs.~\ref{fig:lc_expert} and~\ref{fig:lc_medium} show the training dynamics for the expert and medium datasets, respectively. Since the per-step reward is the negative queueing cost, the episode return and the mean end-to-end packet delay are almost perfectly affinely related in our experiments. Each panel therefore carries a second vertical axis on the right that reads the same curve as mean delay.

During offline pretraining (panels (a)), the greedy evaluation return approaches the average return of the offline dataset (dashed line) within roughly $20$ epochs in both cases: supervised pretraining on only $40$ episodes suffices to imitate the behavior policy, without any environment interaction.

During online fine-tuning (panels (b)), both runs exhibit a transient performance drop over the first ${\sim}25$ iterations. Two factors contribute to this decline. (i) the $\epsilon$-greedy, temporally-committed exploration deliberately injects sustained off-policy power choices for exploration, which depresses the return of the collected episodes (thin curve); and (ii) the newly initialized twin critics are still uninformative while the RL weight $\lambda_{\text{RL}}$ ramps up from zero, so early updates cannot yet exploit the RL gradient. As the critics warm up and the return-weighted replay --- anchored by the best offline trajectories --- concentrates the supervised term on the best behavior observed so far, the evaluation return recovers. Starting from the \emph{expert} pretrained policy (Fig.~\ref{fig:lc_expert}(b)), there is little headroom above the behavior policy, and fine-tuning converges back to the dataset-average level.
Starting from the \emph{medium} pretrained policy (Fig.~\ref{fig:lc_medium}(b)), the evaluation return crosses the offline dataset average at around iteration $70$ and settles clearly above it
: the RL gradient enables the policy to improve \emph{beyond} the quality of its own training data, which pure supervised (offline or online) decision-transformer training cannot do. As a reference point for sample efficiency, pure online MARL in a similar environment requires roughly $500$ episodes of interaction for stable convergence to its optimal performance~\cite{zhang2025multi}, several times the online interaction budget used here. We also note that the stability and coordination mechanisms of Sec.~\ref{subsec: stability coordination} are critical to this success: without them, online fine-tuning in this experiment simply fails to converge.


\subsubsection{QoS performance}


\begin{figure} 
    \centering
    \includegraphics[width=\mywidth]{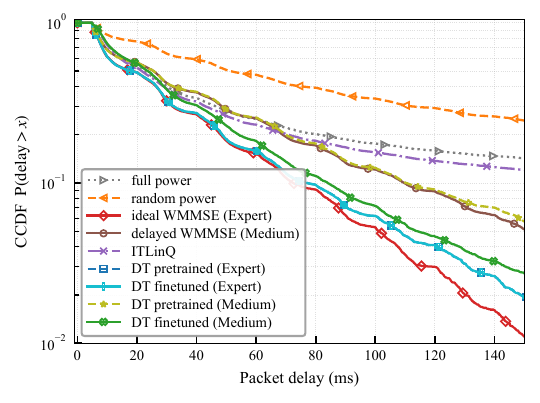}
    \caption{CCDF of packet delays for the 19-device network (log scale), highlighting the delay tail.}
    \label{fig:delay_ccdf}
\end{figure}

\begin{table} 
    \centering
    \caption{Average Packet Delay (ms)}
    \label{tab:delay_stats}
    \begin{tabular}{l|c|c|c}
        \hline
        \textbf{Method} & \textbf{Mean} & \textbf{90th pct.} & \textbf{99th pct.} \\
        \hline
        full power & 96.4 & 273.2 & 1029.2 \\
        random power & 468.2 & 1984.5 & 5362.8 \\
        ITLinQ & 104.1 & 210.2 & 1501.6 \\
        ideal WMMSE (Expert) & 30.5 & 72.4 & 151.7 \\
        delayed WMMSE (Medium) & 44.1 & 109.7 & 252.3 \\
        \hline
        DT pretrained (Expert) & 31.8 & 77.3 & 182.6 \\
        DT fine-tuned (Expert) & 31.8 & 77.3 & 182.6 \\
        DT pretrained (Medium) & 44.8 & 110.3 & 252.0 \\
        DT fine-tuned (Medium) & 37.0 & 84.1 & 221.7 \\
        \hline
    \end{tabular}
\end{table}

Fig.~\ref{fig:delay_ccdf} presents the CCDF of the per-packet delays of all schemes on the common evaluation episode (${\sim}9{,}500$ delivered packets)
, and Table~\ref{tab:delay_stats} reports the corresponding statistics. Three observations follow.

First, offline pretraining faithfully reproduces its behavior policy. The DT pretrained on the expert dataset achieves a mean delay of $31.8$~ms, similar to the genie-aided ideal WMMSE ($30.5$~ms), while operating fully distributedly 
--- in contrast to the global CSI and centralized computation the expert requires. Likewise, the DT pretrained on the medium dataset matches delayed WMMSE almost exactly. 

Second, online fine-tuning is most valuable precisely when the offline data is suboptimal. Fine-tuning the expert-pretrained policy leaves its performance unchanged ($31.8$~ms), consistent with Fig.~\ref{fig:lc_expert}(b): there is essentially no headroom, and the hybrid objective preserves the near-optimal policy. Fine-tuning the medium-pretrained policy, however, reduces the mean delay from $44.8$ to $37.0$~ms --- a $17.5\%$ improvement that also outperforms its own centralized teacher, delayed WMMSE. 
The gains are even more pronounced in the delay tail that dominates user experience: the $90$th-percentile delay drops,  
clearly visible in the CCDF, where the fine-tuned (Medium) curve separates from the overlapping delayed-WMMSE/pretrained pair and moves toward the expert cluster.

Third, all learned policies decisively outperform the practical non-learning references
--- full power, ITLinQ, and random power --- whose delays are markedly larger, with heavy tails exceeding one second at the $99$th percentile (Table~\ref{tab:delay_stats}).

\subsubsection{Discussion}

Beyond the raw delay numbers, two broader implications of this experiment are worth noting. First, the learned DT policy is \emph{fully distributed}: each agent makes its decision using only its own local measurements and limited neighborhood information exchange, with no global CSI and no central coordinator. The per-agent computation and signaling therefore do not grow with the network size, making the solution scalable and feasible for real-world deployment --- in sharp contrast to unscalable centralized WMMSE-type methods. This observation is consistent with the networked-MARL result that policies restricted to $\kappa$-hop neighborhood information can approach globally optimal performance when inter-agent influence decays with graph distance~\cite{qu2022scalable}. In our setting the inter-agent influence is interference, which decays with distance through the path loss, and policies acting on one-hop neighborhood information indeed attain QoS close to that of the centralized, global-information benchmark. Second, the experiment shows that \emph{synthesized} data from a deployment-infeasible algorithm is sufficient to train a deployable policy: ideal WMMSE is unrealizable in a real network, yet the data it generates in simulation distills into a distributed policy that attains nearly the same QoS performance 
using only local information. 

\subsection{Beamforming}

\subsubsection{Simulation setup}

We instantiate the beamforming environment of Sec.~\ref{sec: System Model} on the six-sector cluster of Fig.~\ref{fig:beamforming_deployment}: six APs on a hexagonal ring with $200$~m inter-site distance and inward-facing $120^\circ$ sectors, each serving a single user, with a codebook of $8$ beams of $15^\circ$ width plus the silent action ($9$ discrete actions per agent). We deliberately confine every user to a narrow angular band at the \emph{cell edge} for mobility. 
This stress test places each user in the region of strongest inter-cell interference, keeping the beam decisions of adjacent cells strongly coupled at all times. 

Traffic is Poisson and we report per-packet delay; the remaining parameters are listed in Table~\ref{tab:sim_params}. We compare against three references: \textbf{random} beam selection; a greedy \textbf{max-gain} policy that points each AP directly at its own user (staying silent when its queue is empty) --- a fully distributed heuristic that ignores the interference it inflicts on neighboring cells; and a queue-weighted coordinate-descent \textbf{expert} that performs Gauss--Seidel descent over the discrete joint beam space to maximize the queue-weighted sum-rate using global information --- the analogue of ideal WMMSE over the discrete beam set, and likewise a genie-aided benchmark that is infeasible to deploy. Mirroring the power-allocation experiment, the \emph{expert} offline dataset is generated by this coordinate-descent expert and the \emph{medium} dataset by the max-gain policy, each containing $10$ episodes ($10{,}000$ transitions). Since the problem is much smaller than the $19$-link power-allocation task, we use a lighter decision transformer; 
all other components of the training pipeline of Sec.~\ref{sec: MARL Framework} are unchanged, and all reported evaluations again use the identical channel/traffic/mobility realizations across every stage and baseline.

\subsubsection{Training dynamics and convergence}

\begin{figure*} 
    \centering
    \includegraphics[width=.92\textwidth]{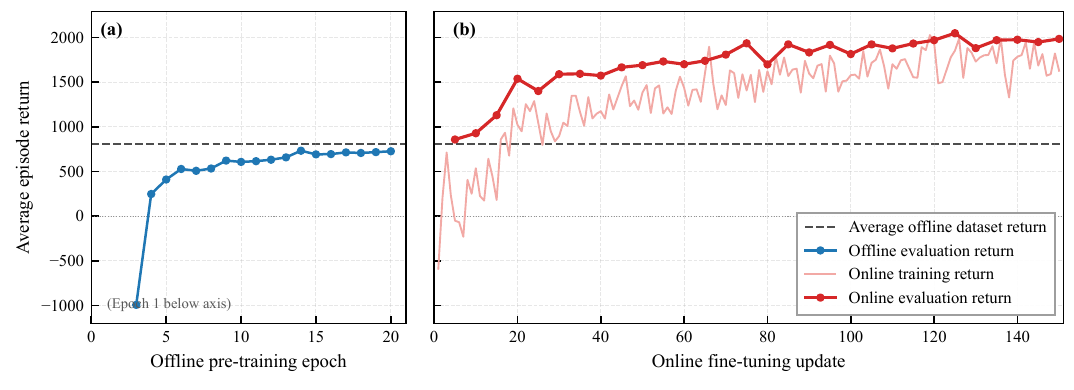}
    \caption{Beamforming learning curves when pretraining on the \emph{medium} (max-gain) dataset: (a) offline pretraining and (b) online fine-tuning. The dashed line marks the average episode return of the offline dataset.}
    \label{fig:lc_beamforming_medium}
\end{figure*}

Fig.~\ref{fig:lc_beamforming_medium} shows the training dynamics for the medium dataset. The delay axis of Figs.~\ref{fig:lc_expert} and~\ref{fig:lc_medium} is omitted here for brevity, as delay remains affinely related to the episode return and the return curve therefore conveys the same information. 
In this experiment the per-slot utility is offset by a constant, $u_k^{(t)} = u_0 - q_k^{(t)}$, which better separates trajectories of different quality in the return-to-go signal without affecting the ranking of policies; it is also why the plotted returns can be positive. 
We also omit the expert-dataset curve: as in the power-allocation experiment (Fig.~\ref{fig:lc_expert}), pretraining imitates the near-optimal expert within a few epochs and fine-tuning leaves it unchanged.

The medium curve is the informative one. In the offline phase (Fig.~\ref{fig:lc_beamforming_medium}(a)), the greedy evaluation return climbs from a large negative value to just below the dataset average within roughly $15$ epochs --- imitation of the max-gain behavior from only $10$ episodes. In the online phase (Fig.~\ref{fig:lc_beamforming_medium}(b)), the exploratory episodes (thin curve) are initially depressed by the committed exploration, but the evaluation return crosses the dataset average within about $20$ iterations and 
settles well above it. The improvement beyond the behavior policy is considerably larger than in the power-allocation experiment (Fig.~\ref{fig:lc_medium}(b)), for a simple reason: in this interference-limited deployment the greedy max-gain policy is much farther from the coordinated optimum than delayed WMMSE is from ideal WMMSE, leaving the RL gradient far more room to exploit.

\subsubsection{QoS performance}

\begin{figure} 
    \centering
    \includegraphics[width=\mywidth]{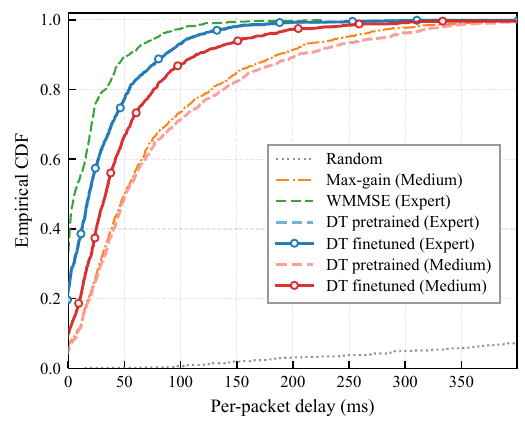}
    \caption{CDF of per-packet delays for the beamforming network. The DT pretrained (Expert) curve is hidden beneath the DT fine-tuned (Expert) curve, with which it exactly coincides.}
    \label{fig:beamforming_delay_cdf}
\end{figure}

\begin{table} 
    \centering
    \caption{Beamforming Packet Delay (ms)}
    \label{tab:beamforming_delay_stats}
    \begin{tabular}{l|c|c|c}
        \hline
        \textbf{Method} & \textbf{Mean} & \textbf{90th pct.} & \textbf{99th pct.} \\
        \hline
        random & 3097.8 & 6241.9 & 7324.9 \\
        max-gain (Medium) & 76.2 & 186.4 & 344.3 \\
        ideal WMMSE (Expert) & 17.6 & 53.8 & 121.5 \\
        \hline
        DT pretrained (Expert) & 31.6 & 86.1 & 184.4 \\
        DT fine-tuned (Expert) & 31.6 & 86.1 & 184.4 \\
        DT pretrained (Medium) & 82.7 & 204.2 & 363.3 \\
        DT fine-tuned (Medium) & 49.5 & 113.4 & 291.4 \\
        \hline
    \end{tabular}
\end{table}

Fig.~\ref{fig:beamforming_delay_cdf} shows the per-packet delay CDF of all schemes on the common evaluation episode (${\sim}1{,}200$ delivered packets), and Table~\ref{tab:beamforming_delay_stats} reports the corresponding statistics. 
The overall picture matches the power-allocation experiment, with a few aspects worth noting:

First, the offline imitation gap on the expert dataset is larger here than in power allocation. We attribute this to the leaner information structure of the beamforming agents: each AP exchanges observations with only its \emph{two} adjacent neighbors, whereas the interference at a cell-edge user is shaped by up to five other sectors. With only partial visibility of the interferers, a distributed policy cannot fully reproduce the decisions of a centralized coordinate-descent expert that sees everything.

Second, online fine-tuning again delivers its largest gains on the medium dataset. The DT pretrained on the medium dataset imitates its max-gain teacher ($82.7$ vs.\ $76.2$~ms; their CDFs in Fig.~\ref{fig:beamforming_delay_cdf} nearly coincide), and fine-tuning then reduces the mean delay by $40\%$ to $49.5$~ms, outperforming its teacher. 
We examined the policies' behavior to find the source of improvement : comparing their actions on the identical evaluation episode, in almost every slot the fine-tuned policy schedules fewer simultaneously transmitting APs than max-gain. Guided by its critic, the fine-tuned policy learns when yielding (going silent or steering away) benefits the neighborhood, and by conceding transmission opportunities to one another the agents jointly reduce interference and raise the realized spectral efficiency, so queued packets drain faster --- exactly the coordination this deployment demands. The random policy, for completeness, is catastrophic: with six sectors interfering over the same region, uncoordinated beams render the network unusable.

Finally, we emphasize the scope of these gains: the advantage of learned coordination is concentrated in the interference-limited regime deliberately constructed here. In additional experiments where users are spread over the full cell area so that inter-cell interference is light, the greedy max-gain policy already attains delay statistics close to the expert's, and there is no meaningful room --- or need --- for a learning-based method. Learned beam coordination is therefore best viewed as a tool for the high-interference operating points (e.g., dense deployments and cell-edge-heavy traffic) rather than as a blanket replacement for simple beam selection.

\subsubsection{Discussion}
Beyond the delay statistics, two broader implications of this experiment are worth noting. First, the beamforming instantiation demonstrates the transferability of the framework: relative to the power-allocation experiment, the action semantics (discrete beam selection vs.\ multi-level power control), the deployment and agent granularity, and the model and data budgets all differ, yet accommodating the new problem required redefining only the per-agent observation, action, and reward, with no algorithmic change to the training pipeline of Sec.~\ref{sec: MARL Framework}.

Second, the experiment characterizes the framework's behavior under restricted information. 
The offline imitation gap to the centralized expert, attributed above to the agents' partial visibility of their interferers, is an honest one that fine-tuning correctly does not overfit away, consistent with the framework's premise that achievable performance is bounded by the local information available to the agents.
\section{Conclusion}
\label{sec:Con}

This paper presented a hybrid offline-online MARL framework that distills the capabilities of model-based optimization methods into practical distributed policies for wireless resource management. In the joint scheduling and power allocation setting as well as the coordinated beamforming setting, the learned policies achieved QoS performance comparable to genie-aided centralized methods using only local measurements and limited neighborhood information exchange. When trained on suboptimal offline data, online fine-tuning further improved performance beyond the data-generating policies.

The framework combines the safety, stability, and sample efficiency of offline pretraining with the adaptability and policy-improvement capability of online RL. Its effectiveness across two settings with different actions, network deployments, agent structures, and training budgets demonstrates its generality. More broadly, the framework enables deployment-infeasible algorithms to provide synthetic training data whose knowledge is distilled into distributed policies. These results position hybrid offline-online MADT as a promising approach to intelligent wireless network management and other distributed multi-agent systems.

\bibliographystyle{IEEEtran}
\bibliography{RLarxiv}

\end{document}